\documentstyle[twoside,fleqn,espcrc1,epsf]{article}
\begin{document}
\newcommand{\ttbs}{\char'134}
\newcommand{\AmS}{{\protect\the\textfont2
  A\kern-.1667em\lower.5ex\hbox{M}\kern-.125emS}}

\hyphenation{author another created financial paper re-commend-ed}

\title{The $SO(5)$ theory of high $T_c$ superconductivity}

\author{Shou-Cheng Zhang\address{Department of Physics, 
        Stanford University, Stanford, CA 94305}
}       

\begin{abstract}
This paper gives a simple introduction to 
the $SO(5)$ theory of high $T_c$ superconductivity. 
Current status and relation to experiments are summarized. 
\end{abstract}

\maketitle

One of the most strikingly universal properties of high $T_c$ 
superconductivity is the close proximity between the superconducting
(SC) and the antiferromagnetic (AF) phases. The AF exchange coupling
$J$ is responsible for the AF phase at half-filling. The same couping
could lead for formation of spin singlets, a prerequesite for 
superconductivity. On the other hand, while the origins of these two
phenomena appear to be related, it is hard to imagine a greater difference
in the physical properties between a insulator and a superconductor.
Is it possible that behind the apparent difference, these
two phases enjoy a deeper and fundamental unity?

Let us take the example of electricity and magnetism in 
vacuum, two of the best
understood phenomena in physics. Before the 18th century, it is widely
believed that these two phenomena are fundamentally distinct. However,
the works of Faraday and Maxwell showed that they are in fact deeply 
related. In the theory of relativity, the electric field $\vec E$ and
the magnetic field $\vec B$ are unified into a electromagnetic field
tensor $F_{\mu\nu}$. The magnetic field $\vec B$ is nothing but the 
electric field $\vec E$ viewed from a rotated coordinate system 
in the four dimensional Minkowski space. Is it possible, that the
two basic phases of the cuperates, AF and SC phases, are in fact 
related to each other by a simple rotation in some higher dimensional 
space?

A recently proposed theory of high $T_c$ superconductivity 
unifies AF and the $d-$wave SC phases and treat them on equal
footing\cite{so5}. 
The AF phase is described by a three dimensional order parameter 
$N_\alpha$, the staggered magnetization. Therefore, it has spin $1$, 
charge $0$ and total momentum $(\pi,\pi)$. On the other hand, a spin
singlet $d$-wave SC phase is described by a complex 
order parameter $\Delta$ 
with two real components, which has spin $0$, charge $\pm 2$ and total 
momentum $0$. The idea of the $SO(5)$ theory 
is to group these five components together into
a object called superspin, $n_a=(Re \Delta, N_x, N_y, N_z, Im \Delta)$
and ask if there exists well-defined rotation operators which can transform
AF into SC and vice versa. Such operators must 
have spin $1$, charge $\pm 2$ and total momentum $(\pi,\pi)$ in order to
patch up the difference between the AF and SC order parameters. These
quantum numbers determine the form of the operator uniquely up
to a form factor. One of them\cite{demler} is given by 
$\pi^\dagger= \sum_{\bf{k}} (\cos k_x - \cos k_y) 
      c_{{\bf k} + {\bf \pi},\uparrow}^\dagger 
      c_{-{\bf k},\uparrow}^\dagger$
Since this operator has spin $1$, one can obviously define three
of them $\pi^\dagger_\alpha$, where $\alpha$ is a spin index. 
This operator also has charge $+2$, one can therefore
define its hermitian
conjugate $\pi_\alpha$ which has charge $-2$. Remarkably, these
operators rotate AF order parameter into the $d$-wave order parameter
\begin{eqnarray}
[\pi^\dagger_\alpha, N_\beta] = i \delta_{\alpha\beta} \Delta^\dagger 
\label{commutator}
\end{eqnarray} 
and vice versa. 
Together with the total spin operators $S_\alpha$ 
and the total charge operator
$Q$, the six $\pi$'s can be identified with the $10$ components of a 
antisymmetric tensor $L_{ab}=-L_{ba}$ in five dimensions, and they
form the generators of the five 
dimensional rotation group, $SO(5)$.
The $SO(5)$ group contains the familiar $SO(3)\times U(1)$ spin
and charge symmetry as a subgroup.

We have thus apparently accomplished the task of unifying AF with SC: 
they are grouped into a five dimensional object and SC is nothing but
AF viewed in some rotated coordinates and vice versa! This construction
looks a bit similar to the unification of $\vec E$ and $\vec B$ by the
Lorentz group. But so far this is only a mathematical fantasy,
we haven't asked if Mother Nature approves the $SO(5)$ symmetry or not.
In the high $T_c$ problem, Mother Nature is very complicated, but
we can check the $SO(5)$ symmetry within some simple models.
One can easily check the $SO(5)$ symmetry by evaluating the commutator
between the Hamiltonian with the $\pi$ operators. Analytical\cite{demler}
and numerical\cite{meixner} works show that that 
the $\pi$ operators are approximate 
eigen-operators of the Hubbard Hamiltonian, in the sense that
\begin{eqnarray}
[{\cal H},\pi^\dagger_\alpha] = \omega_0 \pi^\dagger_\alpha  
\label{eigenoperator}
\end{eqnarray} 
where eigen-frequency $\omega_0$ is of the order of $J$, and proportional
to the number of holes. This relation reminds us of the commutation
relation between the transverse spin components in a magnetic field
and the Zeeman Hamiltonian. Therefore, the $SO(5)$ symmetry is broken
explicitly by the chemical potential, but the pattern of 
explicit symmetry breaking is simple and familiar, and 
therefore easy to handle. In particular, the Casimir operator 
$\sum_{a<b}L_{ab}^2$ (a generalization of the total spin operator 
$\vec S^2$)
of the $SO(5)$ group commutes with the Hamiltonian 
if (\ref{eigenoperator}) is valid. 
The approximate relation (\ref{eigenoperator}) is highly
nontrivial. One could ask if a similar relation would exist for a modified
$\pi$ operator which rotates AF into $s$-wave SC order parameters. The answer
is negative\cite{demler,meixner}. 
Therefore, there is only a approximate symmetry between
AF and $d$-wave SC near half filling.

The ideas of the $SO(5)$ symmetry can now be used to construct a poor
man's model of a high $T_c$ superconductor. The reason that the high
$T_c$ problem appear to be hopelessly complicated is because fermion problems
are hard to deal with. But the recent experimental
discovery of the pseudogap\cite{ong}
gives a ray of hope. At a temperature $T_{MF}$
higher than the true $T_c$, the fermions already appear to be gapped. 
Better yet, these two temperature scales go in opposite direction as
one approaches the AF phase from the SC side. This means that for 
low temperature properties in the underdoped regime, one can forget about
the fermions and concentrate on the collective degrees of freedom, and
map the whole problem into a ``effective magnetic problem" involving the
$SO(5)$ superspin. 

We can define $T_{MF}$ to be the temperature below which the superspin 
acquires a finite magnitude, but its orientation is still not fixed.
Below this temperature, the fermionic degrees of freedom are quenched.
One can always rescale the parameters so that the constraint $n_a^2=1$
is satisfied. We can define superspin vectors locally and end up with
a quantum rotor model with some moment of inertia and gradient coupling. 
At half filling, where the chemical potential is defined to vanish, we assume
that the $SO(5)$ symmetry is broken in such a way that AF phase is favored.
This is similar to a magnetic probelm with ``easy directions". The most 
important question is to ask what happens when the chemical 
potential deviates from zero.  

The chemical potential $\mu$ couples to the charge, one of 
the symmetry generators in the $SO(5)$ theory. It does not couple to 
either AF or SC order parameters directly. Therefore, one's naive 
expectation is that the superspin direction is unaffected. 
However, a close inspection shows that the rotation generators are
not independent of the order parameters. The simplest example of this
kind of relation is the constraint between the total spin $S_\alpha$ 
and the Neel vector $N_\alpha$ of a antiferromagnet: 
$S_\alpha N_\alpha =0$. Neel first derived this result by expressing
$S_\alpha$ and $N_\alpha$ in terms of the sum and the difference
of the sublattice magnetization. However, this orthogonality relation 
has a simple geometric interpretation. The 
$\vec N$ vector is confined to lie on a sphere. The $\vec S$ operator is a 
generator of the rotation of the $\vec N$ vector, therefore it 
has to lie in a plane tangent to the sphere. The Neel orthogonality
relation has a extremely important physical consequence. 
If we apply a uniform magnetic field to a antiferromagnet, it induces 
a total spin along the field direction. The constraint relation 
immediately tells us that the Neel vector has to be perpendicular to the 
field direction. This phenomenon is called a spin-flop transition induced
by a uniform magnetic field.

How does this picture generalize to higher component spins? 
In higher dimensions, the rotation generators can not be described as 
vectors, but can always be described by antisymmetric tensors $L_{ab}$. 
The index $ab$ specify a plane tangent to a sphere traced out by the $n_c$ 
vector, and the generalized orthogonality relation reads 
\begin{eqnarray}
L_{ab} n_c + L_{bc} n_a + L_{ca} n_b =0
\label{constraint}
\end{eqnarray} 
This equation can be proved by expressing the angular momentum operator
in the usual fashion, $L_{ab}=n_a p_b - n_b p_a$, where $p_a$ is the
momentum conjugate of $n_a$, and substituting it 
back into (\ref{constraint}). This simple
equation expresses a geometric property of a hyper-sphere, 
but interpreted in the $SO(5)$ theory, it provides a powerful constraint 
between three most important physical
quantities in the high $T_c$ problem: doping (which is the convention
of \cite{so5} is identified with $L_{15}$), AF ($n_2,n_3,n_4$) 
and SC ($n_1,n_5$). Since only the charge operator $L_{15}$
couples to a external field, namely the chemical potential, the expectation
values of all other generators vanish. The constraint 
equation (\ref{constraint}) immediately 
tells us that at finite doping, the superspin has
vanishing AF components and a finite SC component. Therefore,
while the Neel constraint equation tells us about the physics
of the spin flop, the $SO(5)$ constraint equation (\ref{constraint})
tells us about the physics of the superspin flop,
{\it i.e.} the transition from a AF phase to the SC phase induced 
by the chemical potential. In the $SO(5)$ theory, the mysterious 
transition from AF to SC upon doping is explained by a simple 
geometric property of a sphere is five dimensions!

Once the analogy with the spin flop problem and the superspin flop problem
is realized, one can simply borrow the spin-flop phase diagram to 
construct the phase diagram of the high $T_c$ superconductors in the
temperature and chemical potential plane, see figure 1. When the chemical
potential $\mu$ is less than some critical value, say $\mu_c$, the superspin
prefers to lie in the AF direction. The chemical potential plays the role
similar to the uniform magnetic field in the spin flop problem. Beyond
the critical $\mu_c$, the superspin flops from the AF direction to the SC
plane. When temperature is raised at a constant chemical potential, the
AF state undergoes a second order transition at $T_N$, while the SC state 
undergoes a second order transition at $T_c$. These two second order 
transition lines meet at a bicritical point $T_{bc}$. At this point, the
$SO(5)$ symmetry becomes exact due to critical fluctuations.
Since this point has the most thermal
and quantum fluctuations in the entire phase diagram, 
both $T_c$ and $T_N$ are depressed near $\mu_c$. The $SO(5)$ theory
predicts that both second order lines merge into the first order lines
tangentially, with a behavior close to a square root singularity.
Because of the materials difficulty in the underdoped regime, it is
experimentally unclear if $T_c$ and $T_N$ actually meet at a single point
or are detached from each other. More experimental work in this direction
is clearly desired to test this crucial prediction of the $SO(5)$ theory.

\begin{figure}[h]
\centerline{\epsfysize=4cm
\epsfbox{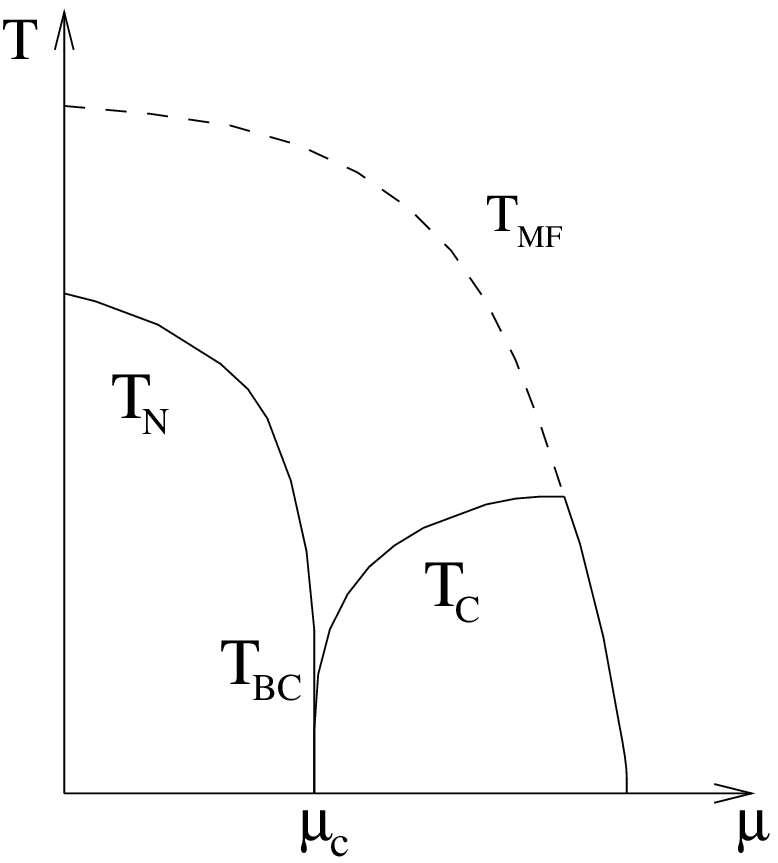}
}
\end{figure}

It is important to note that we plotted the phase diagram as a function
of $\mu$, not doping $x$. Because the density jumps discontinuously across
a first order transition line, the plot as a function of doping
$x$ would contain
a coexistence region. Physics in the region maybe very interesting in itself.
The long range Coulomb interaction could lead to a stripe order of 
alternating AF and SC phases. 
 
So far the most direct evidence of the $SO(5)$ symmetry come from the 
resonant neutron scattering peaks in the $YBCO$ superconductors below
$T_c$\cite{neutron}. These resonances have spin $1$, momentum $(\pi,\pi)$, 
and resolution limited peaks at $41meV$, $33meV$ and $25 meV$ for 
materials with $T_c=92K$, $T_c=67K$ and $T_c=52K$
respectively. The resonance energy scales with $T_c$, but is 
not simply related to the size of
the SC gap,  since recent photoemission experiments show
that the SC gap increases with decreasing doping\cite{shen}.

These resonance peaks have a natural explanation within the $SO(5)$ theory.
In the previous discussions, we argued that beyond a critical chemical 
potential $\mu_c$, the superspin vector lies within the SC plane
$(n_1,n_5)$. However, this notion is a classical one, since the Heisenberg
uncertainty relation does not allow the angle of the superspin to be
sharply defined. In fact, there is zero point motion of the superspin
into the AF directions $(n_2,n_3,n_4)$. What
are the appropriate coordinates describing this zero point motion?
Let us recall the fundamental $SO(5)$ commutation relation
(\ref{commutator}),
which tells us that the $\pi^\dagger_\alpha$ operator rotates AF into SC.
In the SC state, we can approximate the right-hand-side of 
(\ref{commutator}) by a $c$-number expectation value, and this equation
can now be interpreted as the Heisenberg commutation relation between
the canonical momentum $p$ and coordinate $q$ of a harmonic oscillator.
The eigenfrequency of this oscillator can be simply read off from 
equation (\ref{eigenoperator}). This oscillator can be 
naturally identified with the resonant neutron peak observed in the
$YBCO$ superconductors. It has momentum $(\pi,\pi)$, spin $1$ and
a resonance energy which scales with the hole doping. Since the 
harmonic oscillator interpretation crucially depends on the SC order
parameter having a finite expectation value, one would expect the mode
to disappear above $T_c$, which is again consistent with the experiments
in the $T_c=92K$ and $T_c=67K$ superconductors. The $T_c=52K$ material
shows a broad peak even above $T_c$. The situation with this system is
a bit unclear, it could be related to the intrinsic disorder present in 
the $T_c=52K$ material, but we do not have a simple theory for it at this 
moment.
 
A number of other experimental consequences of the $SO(5)$ are currently
being worked out. The $SO(5)$ theory predicts that a SC vortex in the
underdoped regime has a AF core\cite{so5,vortex}. 
Such a configuration is called a ``meron"
in field theory literature. In this configuration, the superspin lies
within the SC plane far away from the vortex core, but it rotates from
the SC plane into the AF sphere as the vortex core is approached in the
radial direction. A vortex lattice with AF core could be detected as
satellite peaks in the elastic neutron scattering experiment. It can also
be detected by muon spin resonance inside the vortex core. Since the AF
vector lies in a plane perpendicular to the applied field, this could give
a distinct signature in the $\mu$SR experiment. The $SO(5)$ theory also
predicts a charge doublet excitation in the AF state, which is the cousin 
of the spin triplet excitation in the SC state. In a AF state, the 
superspin vector lies within the AF sphere, however, its zero point motion
leads to a fluctuation into the SC plane. Similar to the spin triplet
resonance, the charge doublet resonance should only appear below the
Neel temperature $T_N$. Although the phase diagram of the $SO(5)$ 
theory is qualitatively similar to many other alternative theories, it
distinctively predicts a direct first order phase transition from AF 
to SC and a bi-critical point where both $T_N$ and $T_c$ merge. It is 
hard to access the AF/SC transition region experimentally because of 
sample inhomogeneities. This is certainly a major challenge which can hopefully
be resolved experimentally in the near future.

This work is heavily based in the insights gained from the previous
theoretical works in the field. However, due to space limitations,
readers are refered to reference \cite{so5} for detailed discussion
of the relationships. This work is supported in part by the 
NSF under grant numbers DMR-9400372 and DMR-9522915.

\end{document}